# Molecular phases in coupled quantum dots

M. Rontani,[1] S. Amaha,[2] K. Muraki,[3] F. Manghi,[1] E. Molinari,[1] S. Tarucha,[2,3,4] and D. G. Austing[3,5]

[1]INFM–S3 and Dipartimento di Fisica,  Università degli Studi di Modena e Reggio Emilia, Modena, Italy

[2]Department of Physics, University of Tokyo, Bunkyo-ku, Tokyo, Japan

[3]NTT Basic Research Laboratories, NTT Corporation, Atsugi, Kanagawa, Japan

[4]ERATO Mesoscopic Correlation Project, Tokyo, Japan

[5]Institute for Microstructural Sciences, National Research Council of Canada, Ottawa, Canada

We present excitation energy spectra of few-electron vertically coupled quantum dots for strong and intermediate inter-dot coupling. By applying a magnetic field, we induce ground state transitions and identify the corresponding quantum numbers by comparison with few-body calculations. In addition to atomic-like states, we find novel "molecular-like" phases. The *isospin* index characterizes the nature of the bond of the artificial molecule and this we control. Like spin in a single quantum dot, transitions in isospin leading to full polarization are observed with increasing magnetic field.

Whether natural molecules are stable or not depends on the valence electron arangement [1]. For simple homonuclear diatomic molecules like $H_2$, $N_2$, $O_2$, and $F_2$ the relative stability can be understood in terms of the *bond number*, $(N_B - N_{AB})/2$, where $N_B$ ($N_{AB}$) is the number of electrons in the bonding (antibonding) orbitals. A higher bond number means the molecule is more stable. Because the inter-atomic separation is restricted by the competing effects of strong nuclear repulsion and electron-nucleus attraction, these molecules naturally have only one possible very stable ground state (GS) with a unique bond number and (usually) minimum total spin $S_{tot}$. On the other hand, *artificial molecules* (AMs) composed of semiconductor quantum dots (QDs) [2] coupled by both electrostatic Coulomb interaction and quantum mechanical tunneling can overcome this constraint [3,4]. This inter-dot coupling can be controlled in the laboratory, and the additional degree of freedom paves the way for the exploration of new regimes of molecular physics because the nature of the electronic bonds can be tuned.

The high symmetry diatomic AMs we use are made of two vertically coupled QD artificial atoms [Figs. 1(a)-(c)]. The stability of the GS is imposed by external confinement and *all* the orbitals of the parent QDs are valence orbitals (no core electrons). The inter-dot distance $b$ can be varied at growth, and the resulting tunnel coupling between the two QDs residing in the cylindrical mesa gives rise to molecular orbitals spanning the inter-dot barrier which are symmetric (S) or antisymmetric (AS) [Fig. 1(c)]. The *isospin* quantum number, $I=(N_S - N_{AS})/2$ [5], where $N_S$ ($N_{AS}$) is the number of electrons in the S (AS) orbitals, is now the bond number. We induce and clearly identify new transitions leading to changes of $I$ and $S_{tot}$ of the few-electron GSs.

The appearance of new molecular phases in diatomic vertical AMs driven by the change of inter-dot tunneling was recently predicted [3,6-9], but only GS phases in the strong-coupling limit [9,10] and dissociation in the weak-coupling limit [9] were clearly identified. These limiting regimes can be understood in terms of few-electron single-dot physics. In the former, just symmetric GSs [see Fig. 1(c) are filled in a way similar to GSs in a single QD, whilst in the latter just the individual dot states are filled. We now demonstrate the existence of stable states in the intermediate-coupling regime which are distinctly "molecular-like", where the physics is determined by the competition between inter-dot delocalization and electronic correlation [11]. This we accomplish by comparing measured and calculated GS and excited state (ES) spectra of AMs as a function of the magnetic field $B$ for samples with different values of $b$. Analysis of the Configuration-Interaction (CI) wavefunctions reveals that these states have a character mixed between complete delocalization of electrons over both dots and molecular dissociation [12].

Our AM is realized [Figs. 1(a)-(b)] by placing a single gate around a sub-micron cylindrical mesa of diameter $D$ incorporating a triple barrier structure [13]. Current $I_d$ flows through the two QDs, separated by the central barrier of thickness $b$, in response to bias voltage $V_d$ applied between the substrate and top contacts, and voltage on the gate $V_g$. When finite $V_d$ (~1 mV) is applied, the usual Coulomb oscillations broaden into current stripes. We then measure the $N$-electron GS and ES electrochemical potentials within an energy window e$V_d$ [14]. These states appear as steps or peaks within the stripes. It is convenient to show d$I_d$/d$V_g$, in the ($B$, $V_g$) plane on a scale such that the color goes from white ($dI_d/dV_g \leq 0$ S), to blue to red to black ($dI_d/dV_g > 0$ S). The structures are cooled to about 100 mK and $B$ is applied parallel to the current.

We model the AM with the potential $V(\vec{r}) = m^* \omega_0^2 \rho^2 / 2 + V(z)$ - the sum of an in-plane radial harmonic trap and a symmetric square double quantum well (DQW) $V(z)$ along the growth direction $z$, as shown in Fig. 1(c). Here $\vec{\rho} \equiv (x, y)$. Each well is of width $w$ and height $V_0$, and the wells are separated by a barrier of width $b$. $\vec{B}$ is parallel to the $z$-axis [12]. The AM is sufficiently well separated from the top and substrate contact leads by 8 nm ($>b$) tunnel barriers that we can reasonably neglect all details of the interactions with the leads in our model [14, 15]. However, the experimentally determined lateral confinement energy, $\hbar\omega_0(N)$, is an input parameter in the model that mimics the electrode screening as $V_g$ is varied [10-12]. The few-body problem is solved by exact diagonalization of the interacting Hamiltonian in a truncated basis of Slater determinants (CI method) [12]. Symmetry allows us to work within separate subspaces, labeled by the total orbital angular momentum $M$, total spin $S_{tot}$, its projection $S_z$, and parity under spatial inversion. Within each subspace, we find that eigenstates have an almost well defined isospin, even though $I$ is not strictly a true quantum number [5]. The electrochemical potential of the $N$-electron GS or $j$-th ES $\mu_j(N)$ is defined as $\mu_j(N) = E_j(N) - E_0(N-1)$, with $E_j(N)$ the $j$-th ES energy of the $N$-electron system (GS for $j = 0$, ES for $j = 1,2,3,\ldots$).

Figures 1(d) and (e) show calculated and measured $\mu_j(N)$ versus $B$ for $4 \leq N \leq 7$ for $b$=2.5 nm (the most strongly coupled AM). At any given $B$, the lowest edge of each stripe follows $\mu_0(N)$, whilst higher lying lines within the stripe track $\mu_j(N)$.

Upward kinks mark $B$-induced transitions between different GSs- some of which are identified by ▲ [14]. There is good agreement between predicted and observed traces. However, as for single QDs [14], measured ESs are not always as clear as the GSs. Some are very clear (e.g. see 6th stripe), some are weak (marked by dashed lines), and many are unresolved or missing. Relevant electronic configurations are depicted in Fig. 1(d) [14]. Red arrows represent electrons arranged in the S set of Fock-Darwin orbitals (shells of which are separated by energy $\hbar\omega_0$ at 0 T) consistent with the Slater determinant of most weight, as isolated from the CI expansion of the many-body wavefunction. At least up to $N$=7, there is no evidence that any electron goes into an AS orbital [9]. In fact, the observed evolution is strikingly single-dot like [14]. On increasing $B$, electrons in the S orbitals go through a number of transitions that increase $M$, and eventually the spin polarized ($S_{tot}=N/2$) maximum density droplet is reached just beyond 5 T [10]. The non-crossing of the GS and ES for $N$=4 near 0 T [■ in Fig. 1(e)] could be due to small deviations from circular symmetry even though the mesa is circular [16].

From analysis of the CI wavefunctions, we find that electrons occupy almost exclusively the S orbitals. Analogously, in a single QD all electrons are 'frozen' in the ground state of a single quantum well in the $z$ direction. This is due to the energy splitting between S and AS mini-bands ($=\Delta_{SAS}$), which for $b$=2.5 nm is comparable to or larger than the energy separation between orbitals within the same mini-band. At small $b$, electrons are prohibited from the occupation of AS orbitals by the kinetic energy cost $\Delta_{SAS}$. Instead they are spread throughout the whole AM as if it were just a single QD. As $b$ increases, $\Delta_{SAS}$ decreases, and it becomes easier to populate AS orbitals. This can be conveniently described by the *isospin* quantum number $I$. At $b = 2.5$ nm, all GSs present have the maximum allowed value of $I=N/2$.

We focus on the state $(S_{tot}, M, I) = (1/2, 1, 5/2)$ for $N=5$ [marked $\alpha$ in Fig. 1(d)] as a typical configuration attainable for $b=2.5$ nm. In the left column of Fig. 2 we plot the pair correlation function $P_{\sigma\sigma_0}(\vec{\rho}, z; \vec{\rho}_0, z_0)$ for this state ($\sigma = \uparrow, \downarrow$ is the spin) [3]. The upper panels show contour plots of $P_{\uparrow\uparrow}(\vec{\rho}, z; \vec{\rho}_0, z_0)$ in the $x$-$y$ plane, once $z, z_0$ and $\vec{\rho}_0$ are fixed at a maximum of the charge density. The top (middle) panel corresponds to the conditional probability of measuring a spin-$\uparrow$ electron on one dot, dot 1, (the other dot, dot 2,) once a spin-$\uparrow$ electron is fixed on dot 1 ($\bullet$), i.e. $z = z_0 = $ dot 1 ($z = $ dot 2, $z_0 = $ dot 1) [3]. The lower panel shows the "angular correlation". Here we have fixed the modulus $\rho$, and plot $P_{\uparrow\uparrow}(\varphi; \rho, z, \varphi_0, \rho_0, z_0)$ versus the azimuthal angle $\varphi$ ($\varphi_0 = 0$). The black (red) curve refers to electrons on the same (other) dot. The conditional probabilities for electrons on the same dot and on the other dot are almost the same, so we conclude that inter-dot coupling is so strong that all electrons are completely delocalized, i.e. no distinction is possible between the two dots. This holds for all maximum-$I$ states, and illustrates why this AM behaves like a single QD. We can now understand the effect of Coulomb correlation. Firstly, an electron produces its exchange and correlation "hole", so when $\varphi = 0, 2\pi$ the position coincides with the location of the fixed electron, and $P(\varphi) = 0$, due to Pauli exclusion. Secondly, as $\varphi$ is varied, we find two peaks at $2\pi/3$ and $4\pi/3$. The three spin-$\uparrow$ electrons spend most of their time at the vertices of an equilateral triangle to minimize their repulsion [8].

We consider now the effect of decreasing the inter-dot coupling. Figure 3 shows predicted $\mu_j(N)$ and measured stripes in the same way as Figs. 1(d) and (e), but for a sample with a larger inter-dot separation, $b$=3.2 nm. The pattern is qualitatively different, because new phases appear and the AM is no longer single-dot like. We see for $N$ = 4, 5, 7 new molecular states in which at least one electron occupies an AS orbital and consequently $I$ is not always maximal ($<N/2$). Blue arrows represent electrons arranged in the AS set of Fock-Darwin orbitals. Focusing on the 5th stripe near 1 T, in the strong-coupling limit, $N$ = 5 undergoes a GS transition from the state (1/2,1,5/2) to (1/2,4,5/2) [see Fig. 1(d)]. This is due to the "squeezing" effect of $B$ on the wavefunction. At a critical value of $B$ the electrons try to increase their average inter-particle separation by occupying higher angular momentum orbitals. On the other hand, for $b$=3.2 nm, we see that the new molecular GS (1/2,2,3/2), β, appears between the two previous maximum-$I$ GSs [Fig. 3(a)]. This state does exist at $b$=2.5 nm, but it is a highly energetic ES which can not be populated even if $V_d$ is increased so that neighboring stripes touch. Also, see the 4th and 7th stripes in Fig. 3(b) near 0 T, and note how different they are compared to the same stripes in Fig. 1(e).

The pair correlation function of the state (1/2,2,3/2) for $b$=3.2 nm [marked β in Fig. 3(a)] is shown in the center column of Fig. 2. The two dots now are distinguishable, because the probabilities for finding electrons on the same (other) dot are different, and $P(\varphi)$ needs not to be zero anymore when $\varphi = 0,2\pi$ and $z \neq z_0$ (red trace). There is only weak spatial correlation for electrons on different dots, i.e. $P(\varphi)$ is almost flat if $z \neq z_0$, in contrast to the way electrons strongly correlate with each other on the same dot (black trace).

A still weaker coupling regime is explored for $b=$ 4 nm. The predicted spectrum in Fig. 4(a) is very rich, with many GS transitions particularly below 2 T. Here, we find nonmaximal-$I$ states, because now $\Delta_{SAS}$ is sufficiently small that both S and AS orbitals can be readily occupied. Indeed, to the left of the triangles marked $\Delta$, all GSs contain at least one electron in an AS orbital. Focusing again on $\mu_0(5)$, the "hybrid" state (1/2,2,3/2), which occurred as a GS only in a small $B$-range near 1 T for $b=3.2$ nm [β in Fig. 3(a)], is now much more stable and extends over a 1 T range. This is a consequence of the further reduction in $\Delta_{SAS}$. Also, near 0 T, the minimum-$I$ state (1/2,1,1/2), γ, is a $N=5$ GS, whilst near 2 T the familiar maximum-$I$ state (1/2,4,5/2) becomes the GS. The measured spectrum in Fig. 4(b) agrees well with Fig. 4(a), and experimentally, for $N = 4$, 5, 6 and 7 respectively, we can identify the GS transitions $I$=0→1→2, 3/2→5/2, 1→2→3, and 3/2→5/2→7/2. Just as for spin in a single QD [14], these transitions in an AM lead eventually to the full polarization of the isospin ($I=N/2$) at high magnetic field. These $B$-driven $I$-transitions illustrate an alternative way of exploring the phase space, complementary to varying $b$ by growing different samples. Thus, starting from a non-maximal-$I$ GS and switching on $B$, one varies the ratio of $\Delta_{SAS}$ to the energy separation between Fock-Darwin orbitals, and this drives $I$-transitions [7].

To understand the minimum-$I$ condition, in the right column of Fig. 2 we plot the pair correlation function of the (1/2,1,1/2) state at $b=4$ nm [marked γ in Fig. 4(a)]. The constituent dots are now more separated, and electrons tend to develop strong inter-dot correlations leading to a staggered configuration across the central barrier, so the peaks of $P(\varphi)$ for electrons on the same dot and on the other dot are located at alternate positions. We interpret this state as a substantially dissociated molecular state dominated by inter-dot Coulomb interactions. Summarizing, $I$ quantitatively describes the nature of the AM "bond". Maximum $I$ corresponds to a strong "covalent" bond, while minimum $I$ occurs for an "ionic" bond dominated by inter-dot Coulomb correlations. In between these limits, the bond's character is determined by the competition between inter-dot tunneling and electrostatic repulsion.

Our findings show that we are able to scan the phase space of the AM and tune the isospin. The relevant energy scales are $\Delta_{SAS}$ and the separation between Fock-Darwin orbitals. The latter is critically $B$-dependent. If $\Delta_{SAS}$ is relatively large, maximum-$I$ states are favored, while non-maximal-$I$ states appear for smaller $\Delta_{SAS}$. A full description of Coulomb correlation is essential for quantitatively predicting molecular properties. The agreement between measured and calculated spectra is impressive given that real AMs are never perfect [17, 18]. This implies that the effects of unintentional asymmetry or deviations from nominal geometry on the spectra presented are relatively small, i.e. the observed molecular phases are fairly stable [19]. Coupled dots are thus unique laboratories where few-body phases can be easily driven.

Work supported by INFM (SSQI and I. T. Calcolo Parallelo 2003), by MIUR (FIRB Quantum Phases of Ultra-low Electron Density Semiconductor Heterostructures), by the EC (SQID), by the Specially Promoted Research, Grant-In-Aid for Scientific Research, and by DARPA-QUIST program (DAAD 19-01-1-0659). We thank T. Honda for sample processing, and S. Sasaki, K. Ono, Y. Tokura, and G. Goldoni, for useful discussions. Con il contributo del Ministero degli Affari Esteri, Direzione Generale per la

Promozione e la Cooperazione Culturale.

# Appendix A. The Hamiltonian and the Configuration-Interaction method

The single-particle Hamiltonian we use to model the diatomic artificial molecule is $H_0(\vec{r}, s_z) = (-i\hbar\nabla + |e|\vec{A}/c)^2/2m^* + V(\vec{r}) + g^*\mu_B B s_z$, where $V(\vec{r})$ is the single-particle potential discussed in the main text, $\vec{B} = B\hat{z}$ is the magnetic field parallel to the $z$-axis, $\vec{A} = \vec{B}\times\vec{\rho}/2$, $\mu_B$ is the Bohr magneton, $g^*$ is the effective gyromagnetic factor, and $s_z = \pm 1/2$ is the spin. The eigenfunctions of $H_0$ are $\psi_{mni\sigma}(\vec{r}, s_z) = \varphi_{mn}(\vec{\rho})\phi_i(z)\chi_\sigma(s_z)$, where $\varphi_{mn}(\vec{\rho})$ [$m = 0, \pm 1, \pm 2, \ldots$ $n = 0, 1, 2, \ldots$] are the Fock-Darwin orbitals, $\phi_i$ are the symmetric ($i$ = S) and antisymmetric ($i$ = AS) orbitals of the double quantum well, and $\chi_\sigma$ is a two-component spinor ($\sigma = \uparrow, \downarrow$). We neglect higher subbands since the confinement in the $z$-direction is much stronger than in the $x$-$y$ plane. The few-body Hamiltonian $H$ is

$$H = \sum_{i=1}^{N} H_0(\vec{r}_i, s_{zi}) + \sum_{j<i} \frac{e^2}{\kappa_r |\vec{r}_i - \vec{r}_j|} \quad .$$

Note that the total spin $S_{tot}$ is a constant of motion since the symmetry-breaking Zeeman term in $H_0$ is negligible. We take the effective mass $m^* = 0.067m_e$, the dielectric constant $\kappa_r = 12.4$, $g^* = -0.44$, $w = 12$ nm, $V_0 = 250$ meV, and $\hbar\omega_0 = 5.78 N^{-1/4}$ (see also the main text).

The Configuration-Interaction (CI) method consists of directly diagonalizing the Hamiltonian $H$ represented on a basis of Slater determinants (SDs). In principle, if the basis is complete, the solution is exact. However, this is not feasible, so we must truncate this basis and therefore we are forced to optimize it. Our algorithm proceeds in two steps.

(i)     First, we construct the SDs by filling with $N$ electrons a finite number of spin-orbitals $\psi_{mni\sigma}$. The choice of orbitals depends on the value of $\Delta_{SAS}$ (equivalently, $b$), i.e. on how easily AS levels can be populated (see the main text). By trial and error, we find an optimal set of single-particle orbitals for each value of $b$, which are listed in Table I.

(ii)    Now we assume that the Fock space of SDs generated by filling the selected orbitals with $N$ electrons in all possible ways is approximately complete. As an example, in Table II we list the size of the ground state subspaces for $b=3.2$ nm at $B=0$ T for increasing values of $N$. The subspaces are appropriately labelled by the values of the total $M$, parity, and minimum positive value of $S_z$ consistent with $S_{tot}$. The corresponding single-particle orbitals are those shown in Table I. The single-particle basis set is kept fixed for $2 \leq N \leq 7$, and the size of the Fock space increases exponentially with $N$. In order to limit the computational effort, for diagonalization with $N \geq 5$, we introduce an energy cut-off on the average value of $H$ for each single SD. In this way, we are able to limit the maximum linear size of the matrices to $\approx 6 \cdot 10^4$. Once the effective subspace (labelled by $M$, minimum $S_z$, and parity) is selected, our code performs a unitary transformation in order to then rewrite $H$ in a diagonal block form, taking into account the $S_{tot}$ symmetry.

Finally, matrices obtained in this way are handled by the Lanczos package ARPACK [20] run on a SP3 IBM system.

Varying the cut-off, one can estimate the energy accuracy by means of extrapolation to the full size of the subspace. For example, for the subspaces of Table II, we estimate the relative error of the energies used to determine the phase diagram of Fig. 3(a) at $B=0$ T to be 0.001 % for $N=5$, 0.015 % for $N=6$, and 0.86 % for $N=7$. We do not use the extrapolated values since we are interested more in the relative position of the ground and excited states (rather than the absolute energies), and this determines the critical values of $B$ for the ground state transitions, and depends only very weakly on the cut-off energy.

| b=2.5 nm | | | b=3.2 nm | | | b=4.0 nm | | |
|---|---|---|---|---|---|---|---|---|
| (0,0,S) | (0,0,AS) | (1,0,S) | (0,0,S) | (0,0,AS) | (0,1,S) | (0,0,S) | (0,0,AS) | (1,0,S) |
| (1,0,AS) | (-1,0,S) | (-1,0,AS) | (0,1,AS) | (0,2,S) | (1,0,S) | (-1,0,S) | (1,0,AS) | (-1,0,AS) |
| (2,0,S) | (2,0,AS) | (-2,0,S) | (-1,0,S) | (1,0,AS) | (-1,0,AS) | (0,1,S) | (0,1,AS) | (1,1,S) |
| (-2,0,AS) | (0,1,S) | (0,1,AS) | (1,1,S) | (-1,1,S) | (1,1,AS) | (-1,1,S) | (1,1,AS) | (-1,1,AS) |
| (3,0,S) | (-3,0,S) | (3,0,AS) | (-1,1,AS) | (2,0,S) | (-2,0,S) | (2,0,S) | (-2,0,S) | (2,0,AS) |
| (1,1,S) | (-1,1,S) | (2,1,S) | (2,0,AS) | (-2,0,AS) | (2,1,S) | (-2,0,AS) | (2,1,S) | (3,0,S) |
| (-2,1,S) | (3,1,S) | (4,0,S) | (-2,1,S) | (3,0,S) | (-3,0,S) | (3,0,AS) | (4,0,S) | (4,0,AS) |
| (4,0,AS) | (4,1,S) | (5,0,S) | (3,0,AS) | (3,1,S) | (4,0,S) | (5,0,S) | (5,0,AS) | (6,0,S) |
| (5,0,AS) | (0,2,S) | (1,2,S) | (4,1,S) | (4,0,AS) | (5,0,S) | (6,0,AS) | (7,0,S) | (8,0,S) |
| (-1,2,S) | (6,0,S) | (6,0,AS) | (6,0,S) | (7,0,S) | (8,0,S) | (9,0,S) | (10,0,S) | |
| (2,2,S) | | | (9,0,S) | (10,0,S) | | | | |

TABLE I. Single-particle orbitals *(m,n,i)* employed in the construction of Slater determinants for $N$=7 at different inter-dot distances $b$. Each orbital is two-fold spin-degenerate.

| $N$ | size |
|---|---|
| 2 | 43 |
| 3 | 412 |
| 4 | 4442 |
| 5 | 37668 |
| 6 | 223820 |
| 7 | 1165433 |

TABLE II. Dimensions of certain subspaces obtained by filling, with $N$ electrons, the single-particle orbitals listed in Table I for $b=3.2$ nm. The subspaces are labeled by ($M$, $S_z$, parity). In particular, the quantum numbers are (0, 0, gerade) for $N=2$, (0, 1/2, ungerade) for $N=3$, (0, 0, gerade) for $N=4$, (1, 1/2, ungerade) for $N=5$, (0, 0, gerade) for $N=6$, and (0, 1/2, ungerade) for $N=7$. These are the ground state quantum numbers at $B=0$ T. Note that here we choose $S_z \leq S_{tot}$. Also, we use the parity quantum number instead of the isospin, which is refered to in the main text. The former is rigorously a good quantum number (and as such it is implemented in our code), while the latter is a well defined index strictly only in the strong-coupling limit.

# Appendix B. (Iso)spin-blockade and spectral weight $A_{N,N-1}(\omega)$

In a simple picture, (iso)spin-blockade is expected at low temperature when on going from $N-1$ to $N$, $|\Delta S_{tot}| > 1/2$ ($|\Delta I| > 1/2$), and this should lead to a suppression of $I_d$. Provided the transitions are energetically allowed, the current in the $N$-th current stripe of the AM is given essentially by summing all possible paths that an extra electron can tunnel into the ($N-1$)-electron GS, which then becomes the $N$-electron $j$-th state (see e.g. [21]). This tunneling process has the spectral weight

$$A_{N,N-1}(\omega) = \sum_j \left| \left\langle \Psi_{N,j} \left| \sum_{mni\sigma} a^\dagger_{mni\sigma} \right| \Psi_{N-1,0} \right\rangle \right|^2 \delta(E_j(N) - E_0(N-1) - \hbar\omega),$$

where $\left| \Psi_{N,j} \right\rangle$ is the $j$-th $N$-electron state in the second-quantized Fock space and $a^\dagger_{mni\sigma}$ is the fermionic operator which creates an electron in the spin-orbital labeled by the quantum indexes ($m,n,i,\sigma$). Provided the density of states in the contacts have a simple form, the current $I_d$ should effectively be proportional to $A_{N,N-1}(\omega)$ (see e.g. [7,22]).

Figures 1(d), 3(a), and 4(a) in the main text show where different $N-1 \rightarrow N$ transitions can occur but not what the expected current intensities are. Since from our theoretical results one would naively expect several spin- and isospin-blockade regions where the current suppression can occur, it is therefore an interesting issue to check the value of $A_{N,N-1}(\omega)$ in these regions. This is especially true for isospin blockade, since $I$ is only an approximate quantum number, while $A_{N,N-1}(\omega)=0$ in spin-blockade regions by symmetry arguments. In Table III we list all expected spin- and isospin-blockade regions of interest for $N-1$ GS to $N$ GS transitions relevant to Figs. 3(a) and 4(a) for the $b$=3.2 and 4.0 nm AMs (none expected for $b$=2.5 nm), and give computed values of $A_{N,N-1}$. For comparison, we note that after integrating in a small energy interval around a single unblockaded $N–1 \rightarrow N$ transition, typical values attainable for $A_{N,N-1}$ are of the order of unity (the maximum value attainable is exactly one) [7,22].

Inspection of Table III reveals that the isospin-blockade mechanism is expected to be extremely efficient in suppressing certain GS-GS transitions otherwise energetically allowed even though $I$ is only an approximate quantum number.

We see no clear suppression at the expected positions along the bottom edge of the current stripe, or indeed for Coulomb oscillation peaks, in the experimental data. This could be due to the following reasons: i. the proximity of several other unblockaded current-carrying states (channels) near the small regions where blockade is expected that can be populated even at 100 mK and small finite $V_d$ (~1 mV); ii. fluctuations in the density of states of the heavily doped contact regions [21]; and iii. the relaxation time of real spin can be very long [23], but for isospin it is unknown (if it is comparable to the transport time, 1-10 ns, then isospin blockade would be hard to observe). All these reasons complicate the simple picture. Finally we note that our discussion above focused on the simplest possible test of blockade, namely blockade expected for GS-GS transitions. Several GS-ES transitions appear to be weak or absent in the experimental data, but since we could not observe the expected GS-GS regions of blockade, we do not speculate whether (iso)spin-blockade is involved.

| $b$ (nm) | $B$ (T) | $(N\text{-}1,S_{tot},M,I)$ | $(N,S_{tot}^{'},M^{'},I^{'})$ | $A_{N,N\text{-}1}$ | blockade type |
|---|---|---|---|---|---|
| 3.2 | 0.1 | 3,1/2,0,1/2 | 4,1,0,2 | 0.00 | isospin |
| 3.2 | 0.3 | 3,1/2,0,1/2 | 4,0,2,2 | $2.36\ 10^{-3}$ | isospin |
| 3.2 | 0.9 | 5,1/2,2,3/2 | 6,1,3,3 | 0.00 | isospin |
| 3.2 | 1.1 | 5,1/2,2,3/2 | 6,0,6,3 | $5.05\ 10^{-4}$ | isospin |
| 4.0 | 0.0 | 6,1,0,1 | 7,1/2,0,5/2 | 0.00 | isospin |
| 4.0 | 0.1 | 4,0,0,0 | 5,3/2,0,3/2 | 0.00 | spin & isospin |
| 4.0 | 0.13 | 5,3/2,0,3/2 | 6,0,2,1 | 0.00 | spin |
| 4.0 | 0.2 | 6,0,2,1 | 7,1/2,0,5/2 | $4.73\ 10^{-3}$ | isospin |
| 4.0 | 0.6 | 4,0,0,0 | 5,1/2,2,3/2 | $7.90\ 10^{-3}$ | isospin |
| 4.0 | 0.6 | 6,0,2,1 | 7,3/2,3,5/2 | 0.00 | spin & isospin |
| 4.0 | 1.1 | 6,0,2,1 | 7,1/2,6,5/2 | $4.53\ 10^{-4}$ | isospin |
| 4.0 | 1.6 | 3,1/2,0,1/2 | 4,0,2,2 | $4.78\ 10^{-3}$ | isospin |
| 4.0 | 1.7 | 5,1/2,2,3/2 | 6,0,6,3 | $7.05\ 10^{-4}$ | isospin |

TABLE III. (Iso)spin-blockade regions and computed values of the spectral weight $A_{N,N\text{-}1}$. The blockade regions correspond to specific GS-GS transitions of the type $(N\text{-}1,S_{tot},M,I) \rightarrow (N,S_{tot}^{'},M^{'},I^{'})$ with $|\Delta S_{tot}|>1/2$ (spin blockade) and/or $|\Delta I|>1/2$ (isospin blockade). The tabulated regions extend in a (usually narrow) $B$-field range centred approximately at the indicated value of the magnetic field. The spectral weight $A_{N,N\text{-}1}$, computed at the specific value of $B$, is obtained by integration over a small energy interval, $\omega$, centred on a single transition, corresponding to the lowest-energy term of the Zeeman multiplet of the $N$-electron state, namely the state with the maximum $S_z$ allowed.

Note we fix the numerical lower bound of $A_{N,N-1}$ to $2 \times 10^{-5}$. Below this threshold $A_{N,N-1}$ is set to 0.00 in the table.

FIG. 1. (Color) [(a)-(c)] Diagram of mesa containing two coupled QDs, scanning electron micrograph of a typical mesa, and model DQW potential. $b$ = 2.5 nm sample ($D$=0.56 μm). (d) Calculated $B$-dependence of few-electron electrochemical potentials for GSs and first few ESs. Dominant configurations of some states are given [14]. Boxes represent first few Fock-Darwin orbitals (S orbitals only). (e) Corresponding $dI_d/dV_g$ stripes in the $(B, V_g)$ plane, at $V_d$ =1.4 mV. Easily identifiable GS transitions are marked ▲. Some less clear ESs are marked by dashed lines.

FIG. 2. (Color) Calculated pair correlation function $P_{\uparrow\uparrow}(\vec{\rho}, z; \vec{\rho}_0, z_0)$ for $N$ = 5 GSs for $b$ = 2.5, 3.2, 4.0 nm at 0.8, 1.0, and 0.4 T, respectively (α, β, γ). Contour plots on top (middle) row, with a spin-$\uparrow$ electron fixed at $(\vec{\rho}_0, z_0)$ in one dot (●), give the probabilities of finding another spin-$\uparrow$ electron on the same (other) dot in the $x$-$y$ plane. Lengths in units of $l = (\hbar/m^*\omega_0)^{1/2}$, with $l$ =17.1 nm. On the bottom row $P_{\uparrow\uparrow}$ vs. $\varphi$, keeping $\rho = \rho_0$ fixed ($\varphi_0 = 0$). Black (red) curves refer to electrons in the same (other) dot.

FIG. 3. (Color) $b$ = 3.2 nm sample ($D$=0.6 μm). (a) Calculated $B$-dependence of few-electron electrochemical potentials for GSs and first few ESs. (b) Corresponding stripes. Notation is as in Fig. 1. Red (blue) arrows represent electrons in occupied S (AS) orbitals.

FIG. 4. (Color) $b$ = 4.0 nm sample ($D$=0.6 μm). (a) Calculated $B$-dependence of few-electron electrochemical potentials for GSs and first few ESs. (b) Corresponding stripes. Notation is as in Fig. 3. All GSs left (right) of Δ have some (no) electrons in AS orbitals.

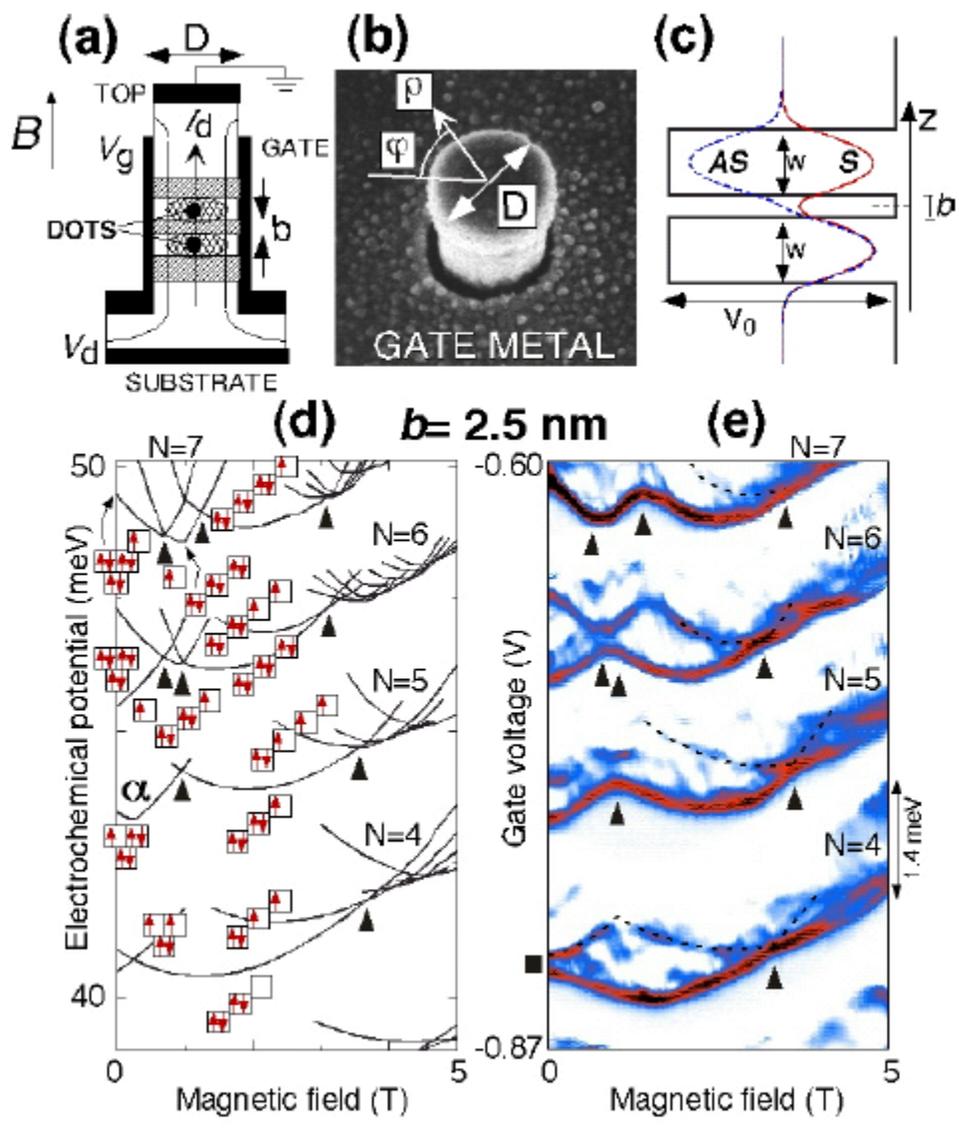

Figure 1 Rontani et al

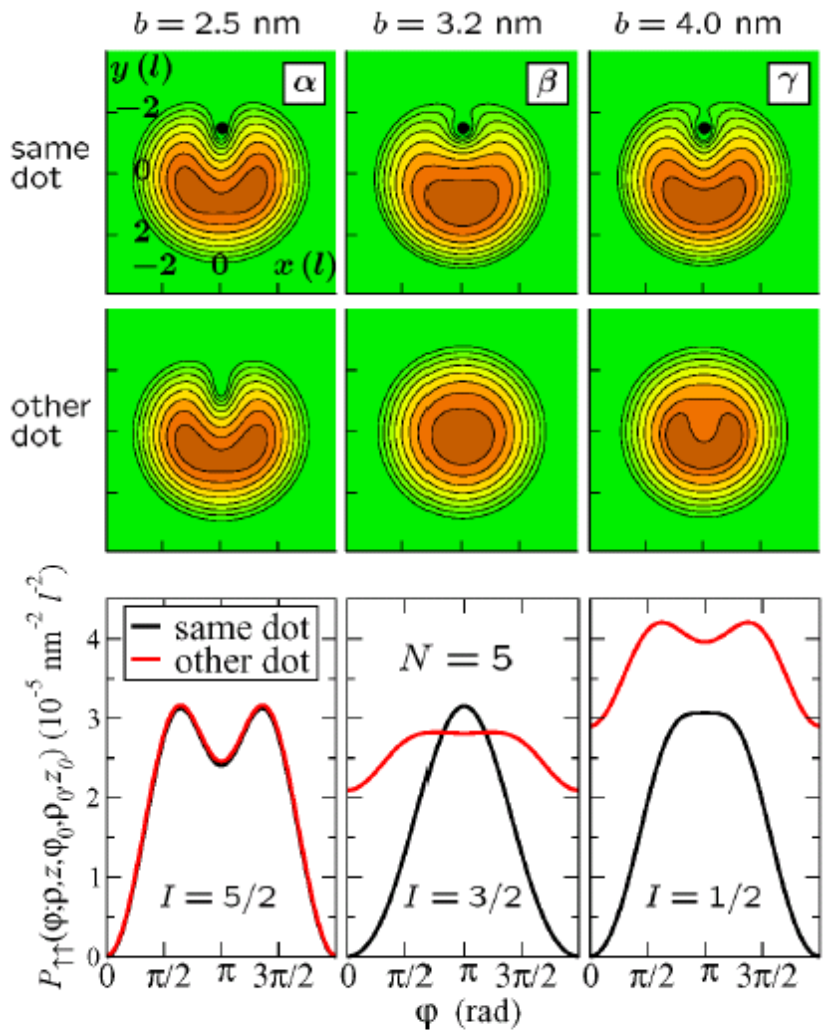

Figure 2 Rontani et al

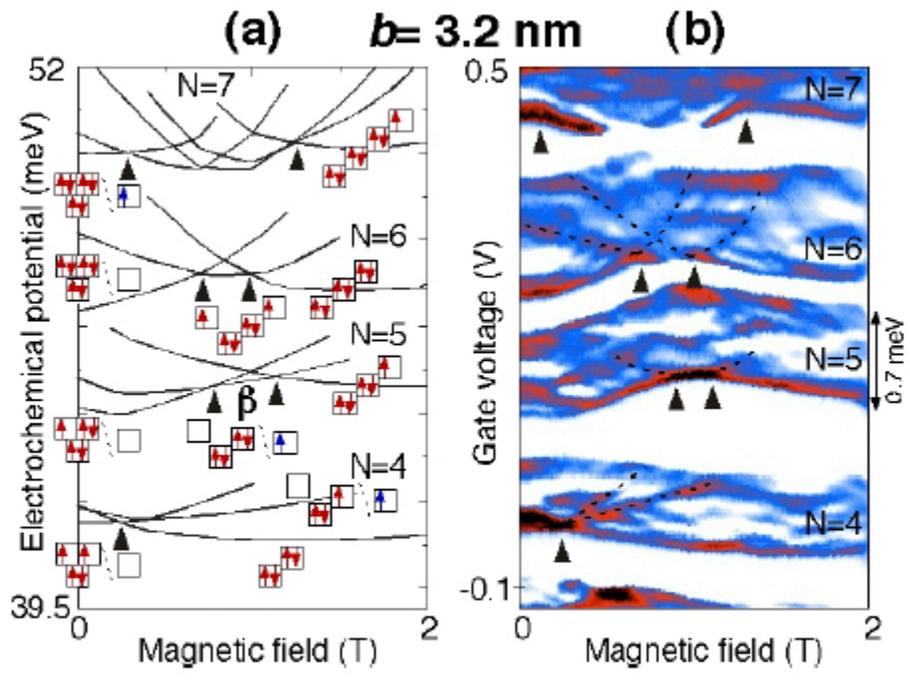

Figure 3 Rontani et al

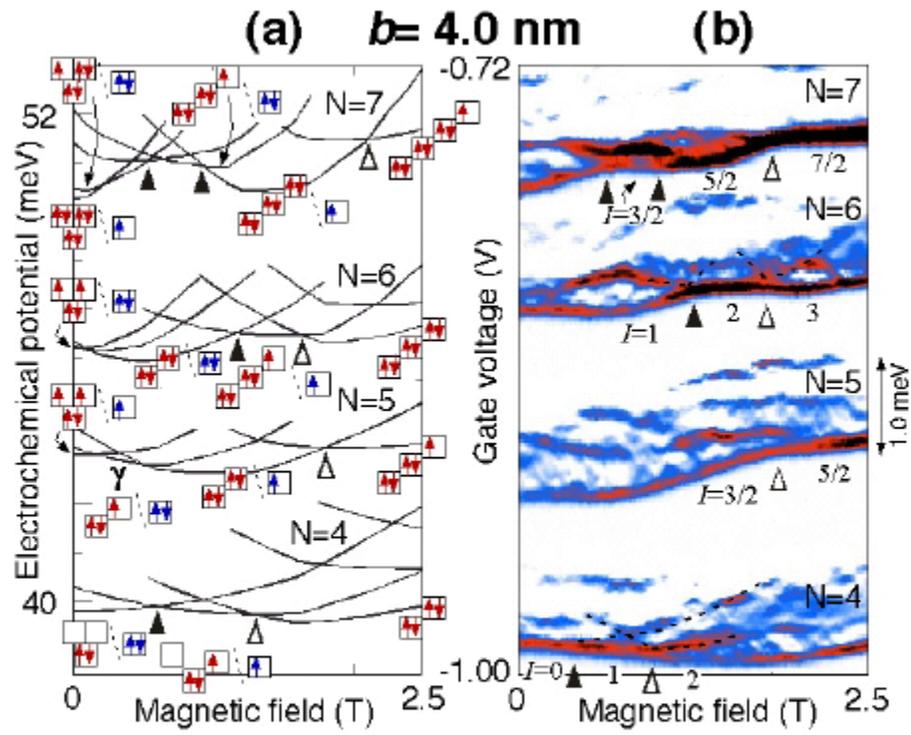

Figure 4 Rontani et al